\documentclass[pre,superscriptaddress,twocolumn,longbibliography]{revtex4-2}
\usepackage[T1]{fontenc}
\usepackage{amsmath}
\usepackage{amssymb}
\usepackage{mathtools}
\usepackage{bbm}
\usepackage{dsfont}
\usepackage[normalem]{ulem}
\usepackage{graphicx}
\usepackage{datetime}
\usepackage{color}
\usepackage{newtxtext,newtxmath}
\usepackage{microtype}
\usepackage{braket}
\usepackage{xr}
\usepackage{natbib}
\usepackage{hyperref}
\hypersetup{%
    bookmarksopen=false,
    bookmarksnumbered=true,
    pdfnewwindow=true,
    unicode=false,
    colorlinks=true,%
    citecolor=blue,
    linkcolor=black,
    urlcolor=blue,
    filecolor=blue
    }

\newcommand{\aDq}[1]{\big\langle\tilde{D}_{#1}\big\rangle}

\newcommand{\Dq}[1]{\tilde{D}_{#1}}
\newcommand{\eps}{\varepsilon}

\DeclareMathOperator{\var}{var}

\newcommand{\vDq}[1]{\var\big(\tilde{D}_{#1}\big)}
\renewcommand{\vec}[1]{\boldsymbol{#1}}

\newcommand{\HSD}{{\mathcal{D}}}
\begin{document}
\newcommand{\figdir}{.}
\newcommand{\usal}{Departamento de F\'isica Fundamental, Universidad de Salamanca, E-37008 Salamanca, Spain}
\newcommand{\iffym}{Instituto Universitario de F\'isica Fundamental y Matem\'aticas (IUFFyM), Universidad de Salamanca, E-37008 Salamanca, Spain}
\title{Characterization of the chaotic phase in the tilted Bose-Hubbard model}%
\author{Pilar Mart\'in Clavero}
\affiliation{\usal}
\author{Alberto Rodr\'iguez}
\email[]{argon@usal.es}
\affiliation{\usal}
\affiliation{\iffym}

\begin{abstract}
The chaotic phase of the tilted Bose-Hubbard model is identified as a function of energy, tilt strength and particle interaction, from the eigenstate structure and the statistical features of the energy spectrum. Our analysis reveals that the chaotic phase of the bare Bose-Hubbard Hamiltonian can actually be enhanced by the presence of a moderate tilt. We further unveil the development and scaling of the chaotic regime from the perspective of a homogeneous density configuration typically used in cold atom experiments, providing a valuable phase diagram for future theoretical and experimental studies of this system.
\end{abstract}
\maketitle
\section{Introduction}
\label{sec:intro}
The dynamical response of a closed quantum system is fundamentally determined by the features of its energy spectrum and the structure of its eigenstates. Notably, when the statistical properties of the spectrum approach those that characterize specific ensembles of random matrices, the time evolution of local observables (i.e., involving a reduced number of the system's degrees of freedom) for appropriate initial states may exhibit a behaviour in agreement with the predictions of statistical mechanics \cite{Deutsch1991,Rigol2008,Srednicki1996,Srednicki1999,Weidenmuller2024b}. Indeed, the standard ensembles of Random Matrix Theory define the benchmark for the certification of quantum chaos \cite{Haake2018,Izrailev1990}, that underlies the potential emergence of ergodic behaviour.

The study of chaotic and non-ergodic regimes in many-particle systems has received a lot of attention in recent years \cite{DAlessio2016,Borgonovi2016,Abanin2019a,Sierant2025}, given the importance of such phases for the possible engineering of a bespoke dynamical behaviour in an otherwise complex system. The identification of the chaotic phase is, e.g., crucial to ensure an adequate performance of current quantum computing architectures \cite{Berke2020,Basilewitsch2023,Borner2023}, and it also proves key for the benchmarking of quantum simulators \cite{Choi2023,Mark2023}.

For interacting systems, the appearance of a chaotic phase typically requires a balance between the characteristic few-particle interaction and the remaining energy scales in the system. A prominent experimental platform for the investigation of this question is given by controlled ensembles of ultracold bosons in optical potentials 
\cite{Kinoshita2006,Trotzky2012b,Cheneau2012,Gring2012,Ronzheimer2013a,Langen2013a,Meinert2014b,Langen2015,Kaufman2016,Choi2016a,Bordia2016,Lukin2018,Rispoli2019,Rubio-Abadal2019,Bohrdt2020,Takasu2020,Leonard2023,Le2023}, which can be used to implement, i.a., the Bose-Hubbard model \cite{Fisher1989,Lewenstein2007,Bloch2008,Cazalilla2011,Krutitsky2016}, a paradigmatic example of a non-integrable quantum many-body system, and extensions thereof. In particular, the tilted Bose-Hubbard Hamiltonian (i.e., in the presence of an additional static force that generates an energy gradient in  the optical lattice modes) has been realized to demonstrate experimentally the rich phenomenology of Bloch oscillations
\cite{BenDahan1996,Peik1997,Peik1997a,Wilkinson1996,Bharucha1997,Madison1997,Madison1999,Anderson1998,Morsch2001,Cristiani2002,Ferrari2006,Kessler2016,Geiger2018a,Masi2021}, including interaction-induced effects 
\cite{Fattori2008,Gustavsson2008,Gustavsson2010,Preiss2015a}, such as the irreversible oscillation decay associated with the emergence of chaotic dynamics \cite{Meinert2014a}. 
The presence of the tilt can be exploited to generate long-range resonant tunneling \cite{Meinert2014b}, as well as to control matter transport by dint of shaking or temporal modulation of the lattice
\cite{Ivanov2008,Sias2008,Alberti2009,Haller2010,Alberti2010,Ma2011,Fujiwara2019}, opening up promising avenues, e.g., to realize anyonic quantum statistics in one dimension \cite{Kwan2023}.

Despite the 
significance of the tilted Bose-Hubbard Hamiltonian,
its chaotic regime \cite{Buchleitner2003,Kolovsky2003c,Kolovsky2004}, which may be associated with unstable Bloch oscillation dynamics \cite{Kolovsky2009,Kolovsky2016a}, has not yet been fully identified.
This work comes to fill this gap, unveiling the location and extension of the chaotic phase as functions of energy and model parameters, that will prove valuable for future theoretical and experimental studies of this system. Interestingly, our analysis reveals that that chaotic phase of the bare Bose-Hubbard model can in fact be enhanced by a moderate tilt. 

The work is organized as follows. An account of the model and its integrable limits is given in Sec.~\ref{sec:model}, while Sec.~\ref{sec:formalism} describes the tools used for the certification of quantum chaos. In Sec.~\ref{sec:Ephase}, an energy-resolved picture of the chaotic phase is presented along with its dependence on tilt and particle interaction. In Sec.~\ref{sec:E0phase}, we analyze the chaotic phase from the perspective of a homogeneous initial density configuration typically used in experiments, and study the scaling toward the thermodynamic limit at fixed particle density in connection with RMT predictions. Finally, Sec.~\ref{sec:conclusions} comprises a summary of the main findings.

\section{Model} 
\label{sec:model}
We study the one-dimensional particle-number conserving Bose-Hubbard Hamiltonian (BHH) \cite{Fisher1989,Lewenstein2007,Bloch2008,Cazalilla2011,Krutitsky2016} of $N$ bosonic particles on $L$ sites in the presence of a static external field that induces a linear tilt of the onsite energies,
\begin{align}
	H =& -J\sum_{j=1}^{L-1}\left(b^\dagger_j b_{j+1} + b_{j+1}^\dagger b_j \right) + \frac{U}{2} \sum_{j=1}^L \hat{n}_j(\hat{n}_j-1)  \notag \\
	 &+ F\sum_{j=1}^L\left(j-\frac{L+1}{2}\right)\hat{n}_j,
	\label{eq:BHH}
\end{align}
where $b_j^{(\dagger)}$ are annihilation (creation) operators associated with Wannier orbitals localized at each lattice site, $\hat{n}_j=b_j^\dagger b_j$, and the lattice constant is set to one. The non-negative parameters $J$, $U$, and $F$, control, respectively, the strengths of nearest-neighbour tunneling, onsite interaction, and tilt, whose effect is chosen to be antisymmetric with respect to the centre of the lattice. Note that the presence of the tilt breaks the parity and translational invariance of $H$. Although the latter may be restored via a time-dependent unitary transformation \cite{Kolovsky2003c}, here, we focus on the characterization of the stationary spectral and eigenvector properties of the system in the presence of hard-wall boundaries. 

Hamiltonian \eqref{eq:BHH} exhibits two integrable limits, where one may find as many independent commuting observables as the system's degrees of freedom, determined by the number $L$ of sites
\cite{PauschThesis}. 
For $J=0$ the many-body eigenstates are trivially given by the Fock states 
\begin{equation}
  \ket{\vec{n}}\equiv \ket{n_1,n_2,\ldots,n_L}
  \label{eq:Fockn}
\end{equation}
in the Wannier basis, $n_j$ being the eigenvalues of the corresponding number operators $\hat{n}_j$. We will refer to this case as the \emph{$J=0$ integrable limit}. In the non-interacting case, $U=0$, the system also admits an analytical solution, the single-particle eigenstates corresponding to Wannier-Stark states \cite{Hartmann2004a}, that are localized in real space with an extension proportional to $J/F$. In the absence of interactions, $H$ would then be diagonal in a Wannier-Stark Fock basis. This case constitutes the $U=0$ or \emph{Wannier-Stark integrable limit}.
\begin{figure}\centering
	\includegraphics[width=.52\columnwidth]{\figdir/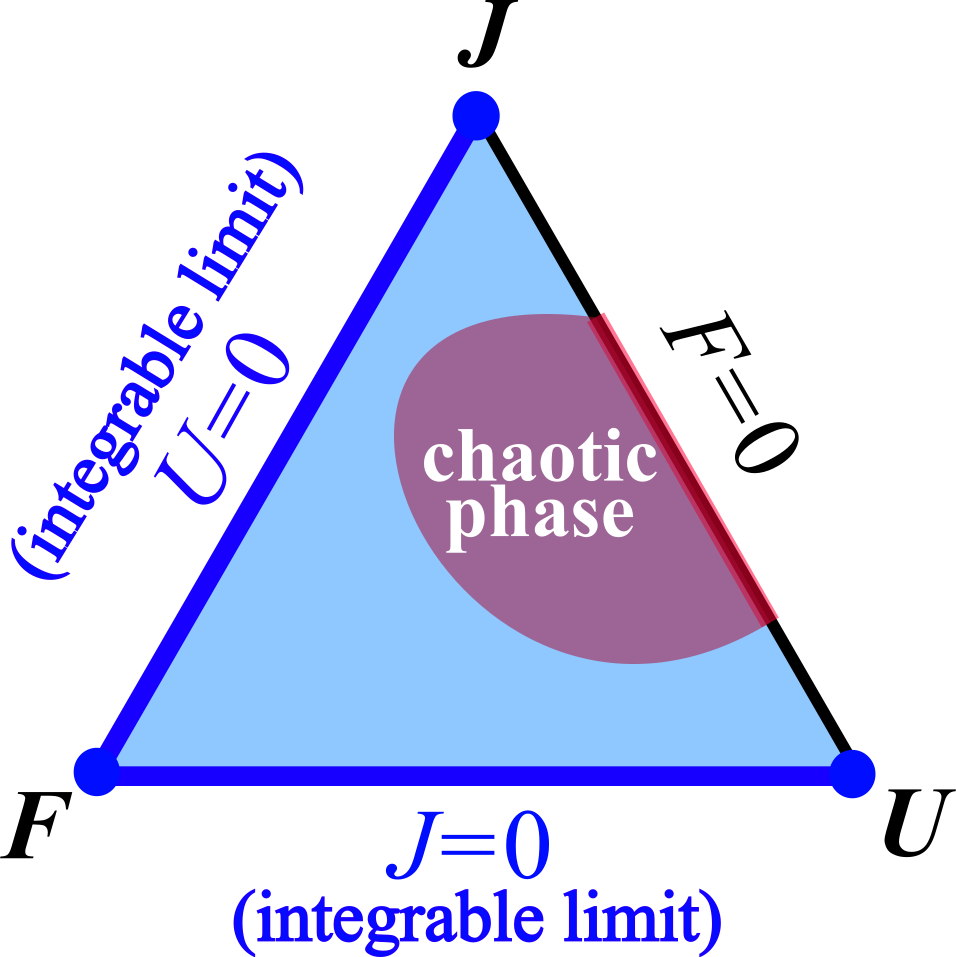}
	\caption{Pictorial representation of the integrable limits and the chaotic phase in the parametric space of the tilted Bose-Hubbard model. The circle nodes (edges) represent configurations where only one term, tunneling, interaction, or tilt, contributes to (is absent from) $H$ [Eq.~\eqref{eq:BHH}]. Dark blue color indicates integrable limits, while light red highlights the chaotic regime.}
	\label{fig:IntDia}
\end{figure}

For $J\neq 0$, $U\neq 0$, and $F=0$, the system becomes non-integrable, and a spectrally chaotic regime emerges in a well defined range of the system parameters \cite{Biroli2010b,Kollath2010,Beugeling2014,Beugeling2015,Beugeling2015c,Dubertrand2016,Beugeling2018,DelaCruz2020,Russomanno2020}. In the absence of tilt, the extension of the chaotic phase \cite{Pausch2020,Pausch2021}, its scaling behaviour with $L$ and $N$ \cite{Pausch2022}, as well as its dynamical manifestations \cite{Duenas2024,Pausch2025}, have been recently characterized in detail. In the case, $J\neq 0$, $U\neq 0$, and $F\neq 0$, the system's non-integrability remains, and the chaotic phase is expected to persist for certain values of $F$ \cite{Buchleitner2003,Kolovsky2003c,Kolovsky2004,Clavero2024}.
The model's integrable limits and chaotic phase are illustrated in the diagram of Fig.~\ref{fig:IntDia}.

\section{Identification of quantum chaos} 
\label{sec:formalism}

The certification of a quantum system's chaotic regime is based upon the benchmarking of certain spectral and eigenvector features against those of the corresponding ensemble, according to symmetry considerations, of random matrix theory (RMT) \cite{bgs84,Bohigas1984a,Berry1985,Guhr1998,Muller2004,Haake2018}. In order to ascertain quantum chaos for $H$ in Eq.~\eqref{eq:BHH}, one must consider the Gaussian orthogonal ensemble (GOE) (real symmetric matrices) of RMT. 

Quantum chaos in the energy spectrum is most conveniently probed via the level spacing ratios $r_n$ \cite{Oganesyan2007,Atas2013c}, 
\begin{equation}
 r_n=\min\left(\frac{s_{n+1}}{s_n}, \frac{s_{n}}{s_{n+1}}\right) \in [0,1],
 \label{eq:rdef}
\end{equation}
where $s_n\equiv E_{n+1}-E_n$ denotes the $n$-th level spacing of the energy spectrum $\{E_n\}$. For GOE matrices, the approximate analytical result for the distribution of this $r$ statistics reads 
\begin{equation}
P_\text{GOE}(r)=\frac{27}{4}\frac{r+r^2}{(1+r+r^2)^{5/2}},
 \label{eq:Prgoe}
\end{equation}
with first moment $\langle r \rangle_\text{GOE}=4-2\sqrt{3}= 0.536$, in agreement with large scale numerics yielding $\langle r \rangle_\text{GOE}= 0.5307$ \cite{Atas2013c}.

The emergence of spectral chaos also entails a fundamental change in the structural properties of the eigenstates, which can be efficiently characterized using finite-size generalized fractal dimensions (GFDs) \cite{Lindinger2019,Rodriguez2011}, 
\begin{equation}
 \Dq{q}=-\frac{1}{q-1} \frac{\ln \sum_{\alpha}|\psi_{\alpha}|^{2q}}{\ln \HSD}, \quad q\in\mathbb{R}^+,
\end{equation}
where $\psi_\alpha$ is the amplitude of the normalized state $\ket{\psi}$ in a given orthonormal basis $\{\ket{\alpha}\}$ of size $\HSD$ in Hilbert space. 
For increasing Hilbert space dimension, 
the GFDs converge to the size-independent fractal dimensions, $D_q\equiv\lim_{\HSD\to\infty} \Dq{q}$, which determine the asymptotic behaviour of the different moments of the distribution of state intensities, $\sum_{\alpha}|\psi_{\alpha}|^{2q}\sim \HSD^{-(q-1)D_q}$, and unveil whether the state is localized ($D_q=0$ for $q\geqslant 1$), ergodic ($D_q=1$ for all $q$), or multifractal ($q$-dependent dimensions $0<D_q<1$) in the chosen basis. Among all the generalized fractal dimensions, the one for $q=1$, 
\begin{equation}
 \Dq{1}=\lim_{q\to1} \Dq{q}=-\frac{1}{\ln\HSD}\sum_\alpha  |\psi_{\alpha}|^2 \ln |\psi_{\alpha}|^2,
 \label{eq:D1}
\end{equation}
determining the scaling of the Shannon information entropy of the state intensities, typically shows weaker finite-size effects than $\Dq{q>1}$. Nonetheless, $\Dq{2}$, related to the scaling of the inverse participation ratio, or $\Dq{\infty}=-\log_\HSD \max_\alpha |\psi_{\alpha}|^2$ which is solely determined by the state's maximum intensity, are also usually considered to analyze the eigenstate structure in Hilbert space \cite{Pausch2020,Pausch2021}. 

For the case of GOE eigenvectors, the behaviour of some generalized fractal dimensions can be analytically exactly or approximately obtained. For instance, the average value and variance of $\Dq{1}$ over GOE  eigenvectors are found to be \cite{Pausch2020}
\begin{align}
 \aDq{1}_\text{GOE} &=1-\frac{2-\gamma-\ln 2}{\ln\HSD} +O\left((\HSD\ln\HSD)^{-1}\right), \label{eq:D1goe} \\
 \vDq{1}_\text{GOE} &= \frac{1}{\ln\HSD^2}\left[\frac{3\pi^2-28}{2\HSD} + O\left(\HSD^{-2}\right)\right].
 \label{eq:vD1goe}
\end{align}
where $\gamma$ is Euler's constant. By construction, the spectrum of GOE matrices is entirely populated by ergodic eigenstates, and therefore $\aDq{q}\to1$ and $\vDq{q}\to 0$ as $\HSD\to\infty$, albeit the convergence toward ergodicity exhibits $q$-dependent finite-size corrections. 

The GFD distributions over close-in-energy eigenstates of a many-body quantum system contains valuable information about the eigenstate structure and its evolution as the system parameters are modified.  
The second moment of such distributions appears to be a particularly sensitive probe to ascertain the emergence of quantum chaos \cite{Pausch2020,Pausch2021,Pausch2022}. 
Additionally, the existence of ground state phase transitions can be efficiently identified from state multifractality in Fock space (which seems to be generic for many-body systems) \cite{Atas2012,Atas2014,Luitz2014,Misguich2017,Lindinger2019}. 

\subsection{Numerical approach}
To determine the emergence of quantum chaos, we calculate numerically the distribution of the level spacing ratios $r_n$ and the generalized fractal dimensions $\Dq{q}$ using exact diagonalization, making use in particular of the PETSc \cite{petsc-user-ref,petsc-efficient,petsc-web-page} and SLEPc \cite{slepc} libraries, which permit an efficient parallelization of the numerical task to be run on High-Performance Computing facilities. A proper strategy using a combination of MPI and OpenMP \cite{Pietracaprina2018} optimize the computational resources needed to obtain a few hundred eigensates in the excitation spectrum of our model for systems with a Hilbert space dimension of $\HSD\lesssim 5\times 10^6$. Additionally, the full spectrum of systems with $\HSD\lesssim 2\times 10^5$ can be tackled using the ScaLAPACK library  \cite{scalapack}. 
\section{Energy-resolved chaotic phase}
\label{sec:Ephase}
In order to ascertain the existence of a chaotic phase in the spectrum of the tilted Bose-Hubbard Hamiltonian [Eq.~\eqref{eq:BHH}], 
we first calculate  the distribution of the level spacing ratios $r_n$ [Eq.~\eqref{eq:rdef}] and the generalized fractal dimension $\Dq{1}$ [Eq.~\eqref{eq:D1}] from the whole spectrum of a system with $N=10$ bosons on $L=10$ sites as functions of the parameters $J$, $U$ and $F$. In this work, we restrict ourselves to lattices at unit filling ($N=L$), and the GFDs are calculated in the onsite Fock basis [Eq.~\eqref{eq:Fockn}].

The behaviour of the above figures of merit in the different regions of the spectrum is analyzed in terms of the rescaled energy 
\begin{equation}
 \eps = \frac{E - E_\text{min}}{E_\text{max} - E_\text{min}},
 \label{eq:epsilon}
\end{equation}%
where $E_\text{min}$ ($E_\text{max}$) is the minimum (maximum) energy for each set of values of the system parameters. Once the bosonic density is fixed, the system spectral and eigenvector properties can be changed by modifying two free parameters. Here, we consider \emph{(i)} $F/U$ and $J/U$, taking $U$ as reference energy, and \emph{(ii)} $F/J$ and $U/J$, after setting the tunneling energy to unity. 

\subsection{Chaotic phase versus $\boldsymbol{J/U}$ and $\boldsymbol{F/U}$}
\begin{figure*}\centering
	\includegraphics[width=.87\textwidth]{\figdir/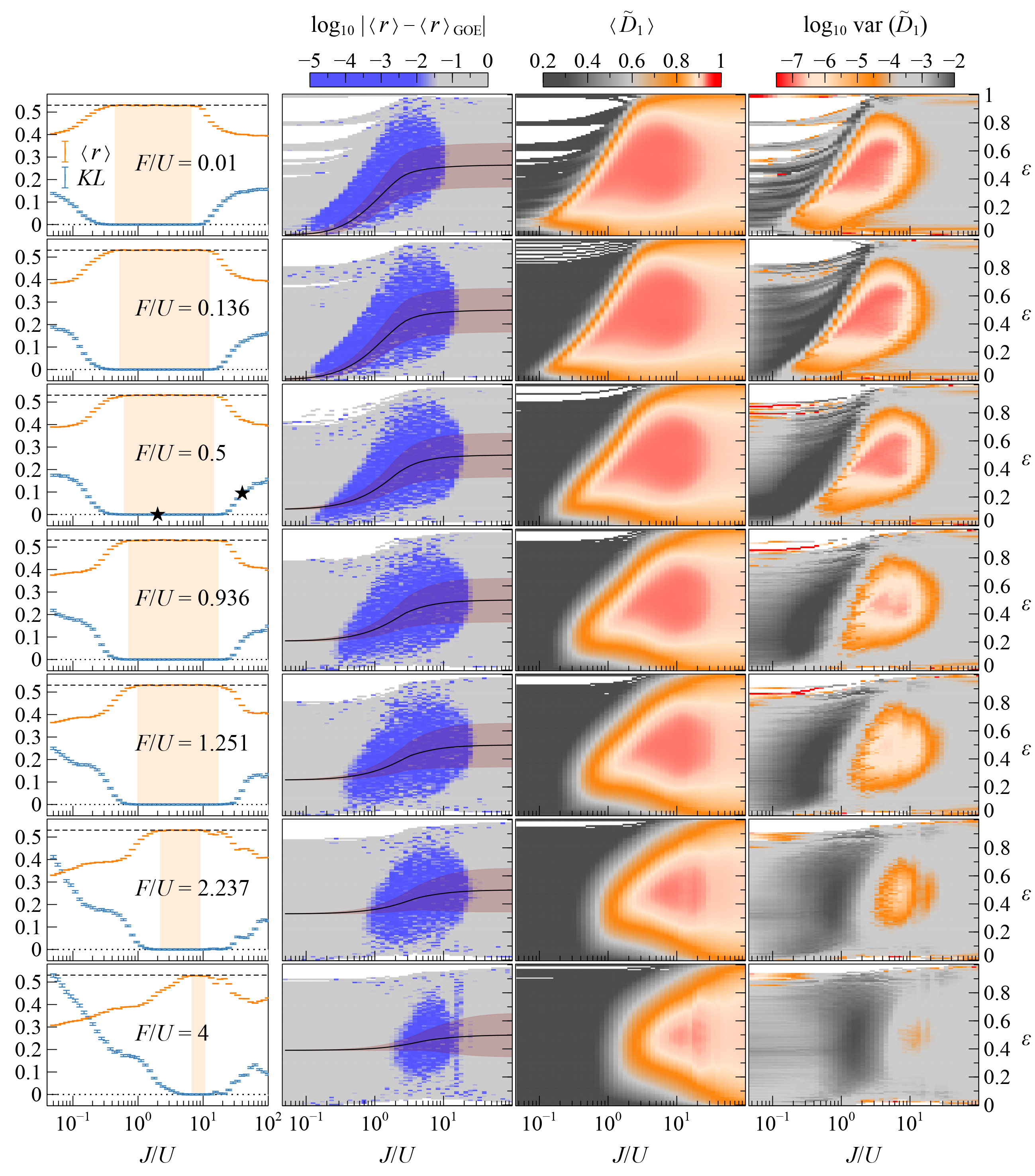}
	\caption{Evolution of the chaotic phase for $L=N=10$ ($\HSD=92\,378$) as a function of $J/U\in[0.05,100]$ (50 equally spaced values in log scale) for increasing values of $F/U$ (from top to bottom) as indicated. The left column shows the mean level spacing ratio $\langle r \rangle$ and the Kullback-Leibler divergence ($KL$) [Eq.\eqref{eq:KL}] obtained from the inner 80\% of the spectrum, where the horizontal dashed line marks the value $\langle r\rangle_\text{GOE}=0.5307$. The shaded areas in the left column highlight the parametric ranges where $\langle r\rangle$ agrees with RMT up to a tolerance of 1\%. Star symbols indicate the parameter values considered in Fig.~\ref{fig:KLdistributions}. Density plots display the deviation $|\langle r\rangle -\langle r\rangle_\text{GOE}|$, the mean fractal dimension $\aDq{1}$, and the variance $\vDq{1}$, resolved in terms of the scaled energy $\varepsilon$ [Eq.~\eqref{eq:epsilon}]. Black solid lines and the accompanying shaded region mark the energy trajectories of the homogeneous Fock state $\ket{\vec{1}}$ [Eq.~\eqref{eq:Fock1}] and its energy width $\pm\sigma_{\ket{\vec{1}}}$ [Eq.~\eqref{eq:sigma1}], respectively.}
	\label{fig:OverviewU1}
\end{figure*}

The energy-resolved mean value of the level spacing ratios, $\langle r \rangle$, the mean fractal dimension $\aDq{1}$, and the corresponding variance $\vDq{1}$ as functions of $J/U$ for different values of $F/U$ are shown in the density plots (three rightmost columns) of Fig.~\ref{fig:OverviewU1}. These three quantities are calculated after dividing the rescaled energy axis into 100 equal bins, considering all the energy levels and eigenstates within each $\eps$ bin. Additionally, the left column in Fig.~\ref{fig:OverviewU1} provides an energy-integrated picture of the properties of the $r$-statistics distribution by displaying $\langle r \rangle$ and the Kullback-Leibler divergence $KL(P,P_\text{GOE})$ for the inner $80\%$ of the energy spectrum. The Kullback-Leibler divergence \cite{Kullback1951} provides a comparison between the model's and RMT distribution for the level spacing ratio,
\begin{equation}
 KL(P,P_\text{GOE})=\int_0^1 \textrm{d}r\, P(r) \ln\left(\frac{P(r)}{P_\text{GOE}(r)}\right),
 \label{eq:KL}
\end{equation} 
and is here estimated by considering the discretized version of the latter equation for the finite difference $\Delta r=0.04$. 

We monitor the evolution of the chaotic phase emerging in the excitation spectrum by tuning $J/U$ for 14 different values of $F/U\in[0.01,4]$, but only show in Fig.~\ref{fig:OverviewU1} the results for 7 tilt values that capture the main features of the $F/U$ dependence. 

For very low tilt, $F=0.01$ (top row in Fig.~\ref{fig:OverviewU1}), one consistently recovers the picture in the absence of external field \cite{Pausch2020}: The chaotic phase, as revealed by the $\langle r\rangle$ density plot, distinctively appears in the range $0.1\lesssim J/U \lesssim10$, albeit with a marked energy dependent onset that shifts toward higher $J/U$ values the higher the energy, giving rise to a characteristic tilted structure. This spectrally chaotic regime is also unveiled by $\aDq{1}$ and $\vDq{1}$, and correlates unambiguously with the appearance of a region populated by extended states in Fock space ($\aDq{1}\to 1$) that tend to exhibit the same delocalized structure [$\vDq{1}\to0$] (the scaling of GFD features with Hilbert space dimension is discussed later on in Sec.~\ref{sec:E0phase}). In particular, note how $\vDq{1}$ undergoes a very pronounced drop by several orders of magnitude in the chaotic regime, making this figure of merit an extremely sensitive probe of quantum chaos. The corresponding energy-integrated picture of the chaotic phase can also be distinctively identified in the top left panel, not only by the mean level spacing ratio, but also by the behaviour of the entire $P(r)$ distribution, as indicated by the vanishing of the Kullback-Leibler divergence. 

\begin{figure}\centering
	\includegraphics[width=\columnwidth]{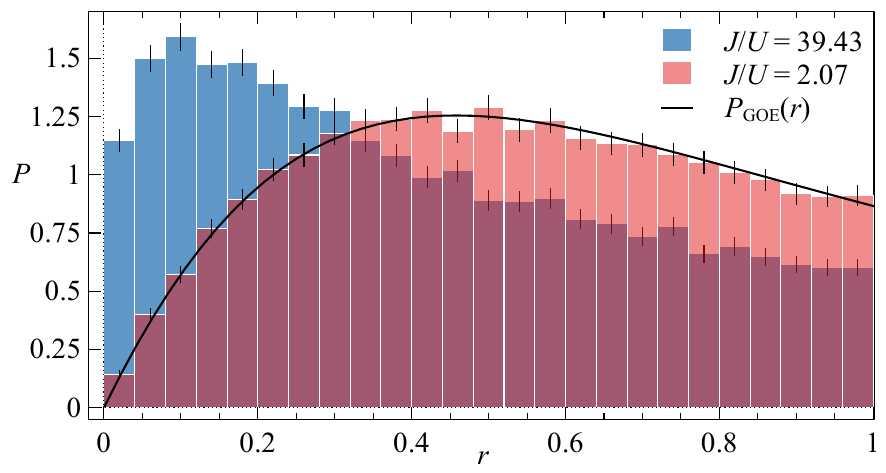}
	\caption{Numerical distributions of the $r$ statistics [Eq.~\eqref{eq:rdef}] obtained from the inner 80\% of the spectrum for $L=N=10$ ($\HSD=92\,378$), $F/U=0.5$, $J/U=2.07$ and $J/U=39.43$ (points indicated by star symbols in Fig.\ref{fig:OverviewU1}), using a binning of $\Delta r=0.04$. The solid line corresponds to the analytical GOE result given in Eq.~\eqref{eq:Prgoe}.}
	\label{fig:KLdistributions}
\end{figure}

As the tilt strength is increased, the chaotic phase seems to remain fairly stable for $F/U\lesssim 0.9$. In order to check the consistency of the observed $KL(P,P_\text{GOE})$ values for non-vanishing tilts, Fig.~\ref{fig:KLdistributions} shows the $P(r)$ distributions for $F/U=0.5$ and two values of $J/U$ (indicated by star symbols in Fig.~\ref{fig:OverviewU1}), confirming the agreement with RMT in the chaotic phase. 

For $F/U\gtrsim 1$, the onset of chaos starts to recede toward larger $J/U$ values and the overall shape of the chaotic phase shrinks. 
For $F/U=4$, while some traces of a narrow spectrally chaotic domain remain in the energy spectrum, the fingerprint of eigenstate ergodicity in $\vDq{1}$ is almost entirely gone. Naturally, in the limit $F/U\to\infty$ the system must be  integrable for any value of $J/U$: The eigenstates of the system would evolve from being Fock states in the onsite basis ($J/U\to 0$) to becoming Fock state in the Wannier-Stark basis ($J/U\to\infty$). 

\begin{figure}\centering
	\includegraphics[width=.95\columnwidth]{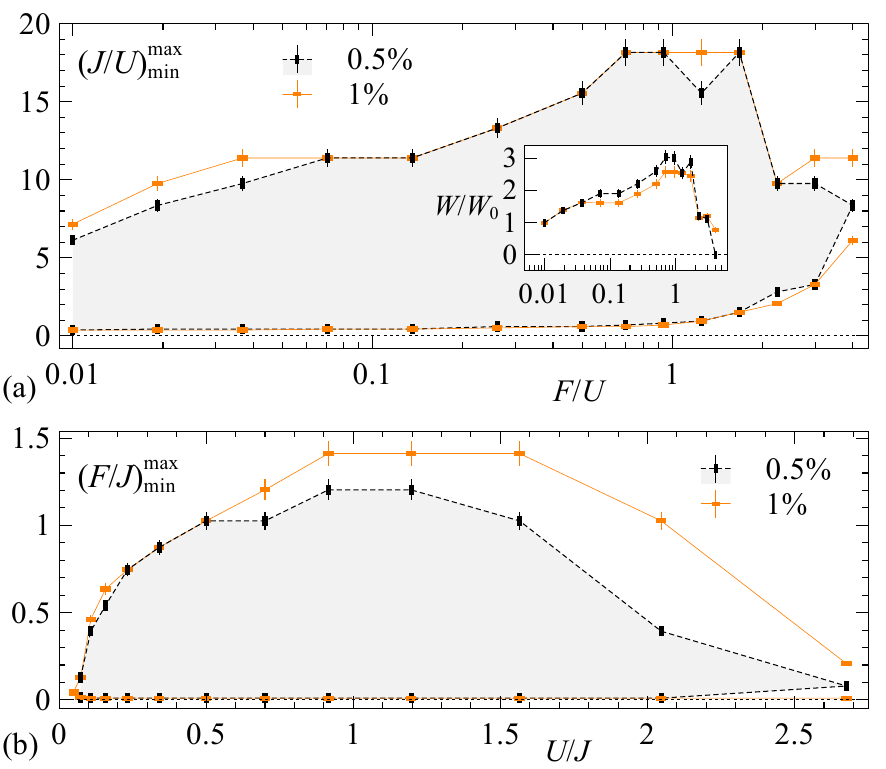}
	\caption{Parametric width of the chaotic phase for $L=N=10$ determined from the comparison of $\langle r\rangle$ calculated over the inner 80\% of the entire spectrum and  $\langle r\rangle_\text{GOE}$. Panel (a) shows the max (upper trajectories) and min values (lower trajectories) of the $J/U$ interval within which  $\langle r\rangle$ agrees with RMT up to a tolerance of 0.5\% (black) or 1\% (orange) as functions of $F/U$ (cf.~left column in Fig.~\ref{fig:OverviewU1}). The shaded area highlights the evolution of the width $W$ of the chaotic phase, which is shown in the inset normalized to its value $W_0$ when $F\to0$. Similarly, panel (b) shows the evolution of the width of the chaotic phase in terms of $F/J$ as a function of $U/J$ (cf.~left column in Fig.~\ref{fig:OverviewJ1}).}
	\label{fig:ChaoticWidth}
\end{figure}
Remarkably, visual inspection of the evolution versus $F/U$ suggests that the extension of the chaotic domain, as compared to the untilted case, may actually grow for moderate $F/U$ values. This behaviour is clearly confirmed by $\langle r\rangle$ and $KL$ in the energy-integrated analysis. The shaded regions in the left column of Fig.~\ref{fig:OverviewU1} highlight the width of the chaotic phase, defined as the $J/U$ parametric region where the mean level spacing ratio deviates less than $1\%$ from the GOE benchmark. The dependence of the limits and width of those regions on $F/U$ is shown in Fig.~\ref{fig:ChaoticWidth}(a), where one observes that the extension of the chaotic phase is enhanced by a factor of three at $F/U\approx 0.8-0.9$ with respect to the untilted case. While the lower limit of the chaotic region remains almost insensitive to the tilt for $F/U\lesssim 1$, the upper limit grows on average linearly with $F/U$, saturating around $F/U\approx 0.8$. For $F/U>1$ the upper limit decays abruptly, the lower limit rises, and the chaotic region shrinks. This behaviour remains qualitatively the same if the tolerance threshold to compare against RMT is reduced to $0.5\%$, as seen in Fig.~\ref{fig:ChaoticWidth}(a).

The presence of a tilt then delays the appearance of the regularity in the spectrum induced by the non-interacting integrable point ($J/U\to\infty$) as long as the interaction term is larger than or comparable to the tilt ($F/U\lesssim 1$), hence enhancing the chaotic phase. One may argue that since in the non-interacting limit the eigenstates in the onsite Fock basis, although not ergodic, exhibit a high degree of delocalization (they are Fock states in the momentum basis), as $J/U$ is reduced from $J/U=\infty$, the presence of a small tilt aids the interaction-induced level mixing to turn such nearly delocalized states into ergodic states, i.e., as $J/U$ decreases the chaotic phase develops earlier than in the absence of tilt. On the other hand, if the tilt dominates over the interaction, the system would start to be pulled toward the Wannier-Stark integrable limit for moderate values of $J/U$, and thus the dwindling chaotic phase observed for $F/U>1$. 

\begin{figure*}\centering
	\includegraphics[width=.87\textwidth]{\figdir/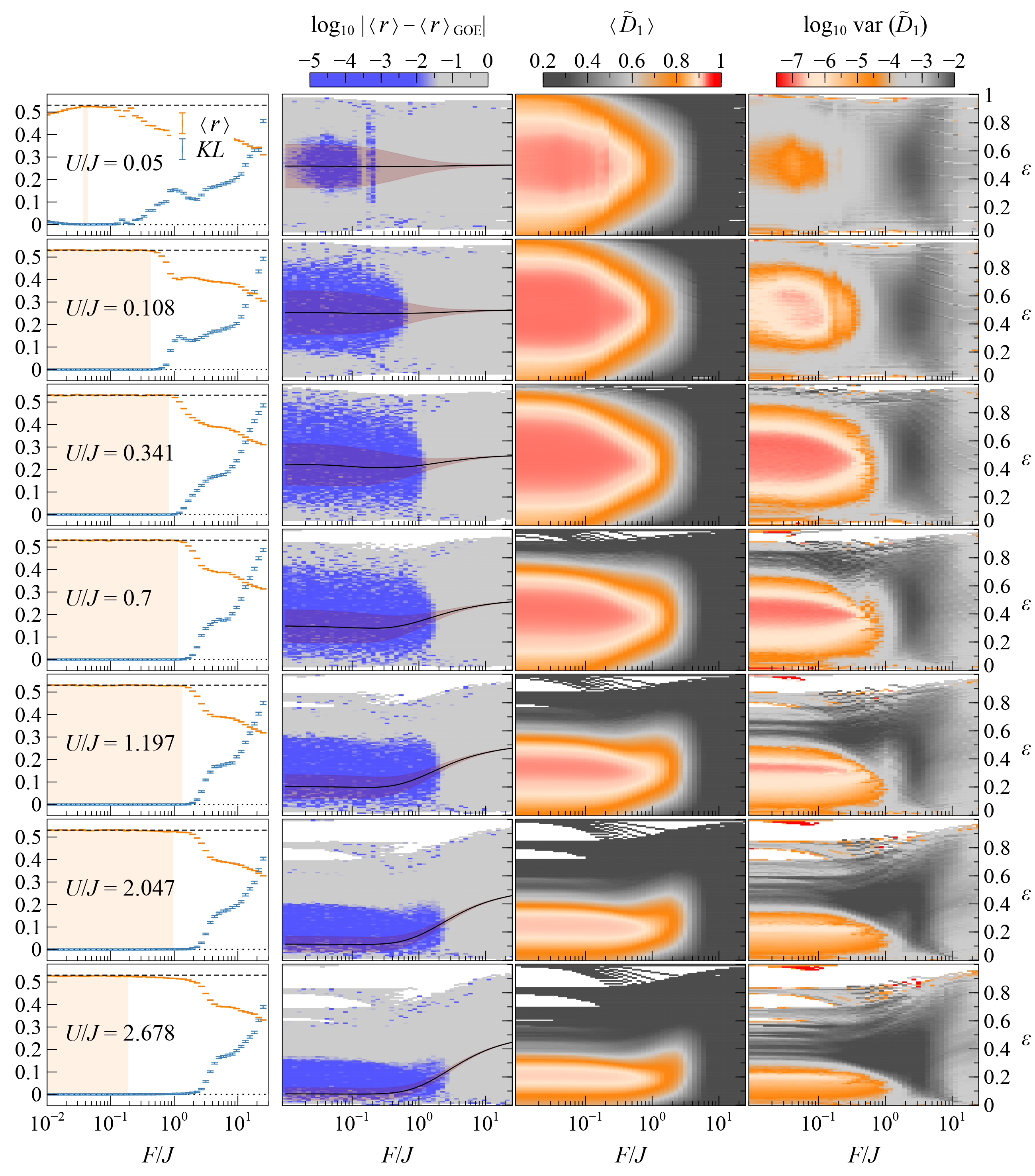}
	\caption{Evolution of the chaotic phase for $L=N=10$ ($\HSD=92\,378$) as a function of $F/J\in[0.01,25]$ (50 equally spaced values in log scale) for increasing values of $U/J$ (from top to bottom) as indicated. The left column shows the mean level spacing ratio $\langle r \rangle$ and the Kullback-Leibler divergence ($KL$) [Eq.\eqref{eq:KL}] obtained from the inner 80\% of the entire spectrum, where the horizontal dashed line marks the value $\langle r\rangle_\text{GOE}=0.5307$. The shaded areas in the left column highlight the parametric ranges where $\langle r\rangle$ agrees with RMT up to a tolerance of 1\%. Density plots display the deviation $|\langle r\rangle -\langle r\rangle_\text{GOE}|$, the mean fractal dimension $\aDq{1}$, and the variance $\vDq{1}$, resolved in terms of the scaled energy $\varepsilon$ [Eq.~\eqref{eq:epsilon}]. Black solid lines and the accompanying shaded region mark the energy trajectories of the homogeneous Fock state $\ket{\vec{1}}$ [Eq.~\eqref{eq:Fock1}] and its energy width $\pm\sigma_{\ket{\vec{1}}}$ [Eq.~\eqref{eq:sigma1}], respectively.}
	\label{fig:OverviewJ1}
\end{figure*}

\subsection{Chaotic phase versus $\boldsymbol{F/J}$ and $\boldsymbol{U/J}$}
In order to get a complete perspective of the chaotic regime, we also study its dependence on tilt and interaction strength when choosing the tunneling parameter $J$ as the reference energy. Figure \ref{fig:OverviewJ1} shows the emergence of the chaotic phase as a function of $F/J$ for several values of $U/J$, using the same figures of merit and layout as in Fig.~\ref{fig:OverviewU1}.

For very weak interactions ($U/J=0.05$), the chaotic phase is absent for $F/J=0$ but makes a timid appearance for small tilt strengths, centered around $F/J\approx 0.05$, confirming that the tilt enhances the interaction-induced level mixing close to the non-interacting limit, as we discussed earlier. For stronger interactions the chaotic phase appears well developed in the region $F/J\lesssim 1.5$ for $0.1<U/J\lesssim 3$, to eventually disappear as $U/J\to\infty$. Independently of the interaction strength, once $F$ exceeds $J$, nearest-neighbour tunneling becomes markedly off-resonant and gets rather  suppressed, hence the system approaches the influence of the $J=0$ integrable limit. 

It is also noticeable that for increasing interaction strength the chaotic phase is pushed down toward lower rescaled energies: Higher $U/J$ promotes a tail of states with high energies separated by pronounced energy gaps in the density of states, and thus the bulk of the spectrum shifts to lower $\eps$ \cite{Pausch2021}. This is also the cause for the elongation of the chaotic phase toward lower rescaled energies for small $J/U$ and low tilts observed in Fig.~\ref{fig:OverviewU1}. 

As done in the previous analysis, we study the dependence of the $F/J$ parametric width of the chaotic phase on $U/J$ from the energy-integrated behaviour of the mean level spacing ratio (see shaded areas in the left column of Fig.~\ref{fig:OverviewJ1}). The evolution of the  $F/J$ limits of the chaotic regime versus interaction strength is shown in Fig.~\ref{fig:ChaoticWidth}(b). The upper limit can be seen to evolve on average nearly quadratically with $U/J$, while the lower limit remains essentially fixed at $F/J\approx 0$. In terms of the range of values for the tilt strength, the extension of the chaotic regime is maximized for $U/J\approx 1$ and vanishes for $U/J>3$.

\section{Emergence of chaos from the perspective of a homogeneous density configuration}
\label{sec:E0phase}
\begin{figure*}\centering
	\includegraphics[width=.85\textwidth]{\figdir/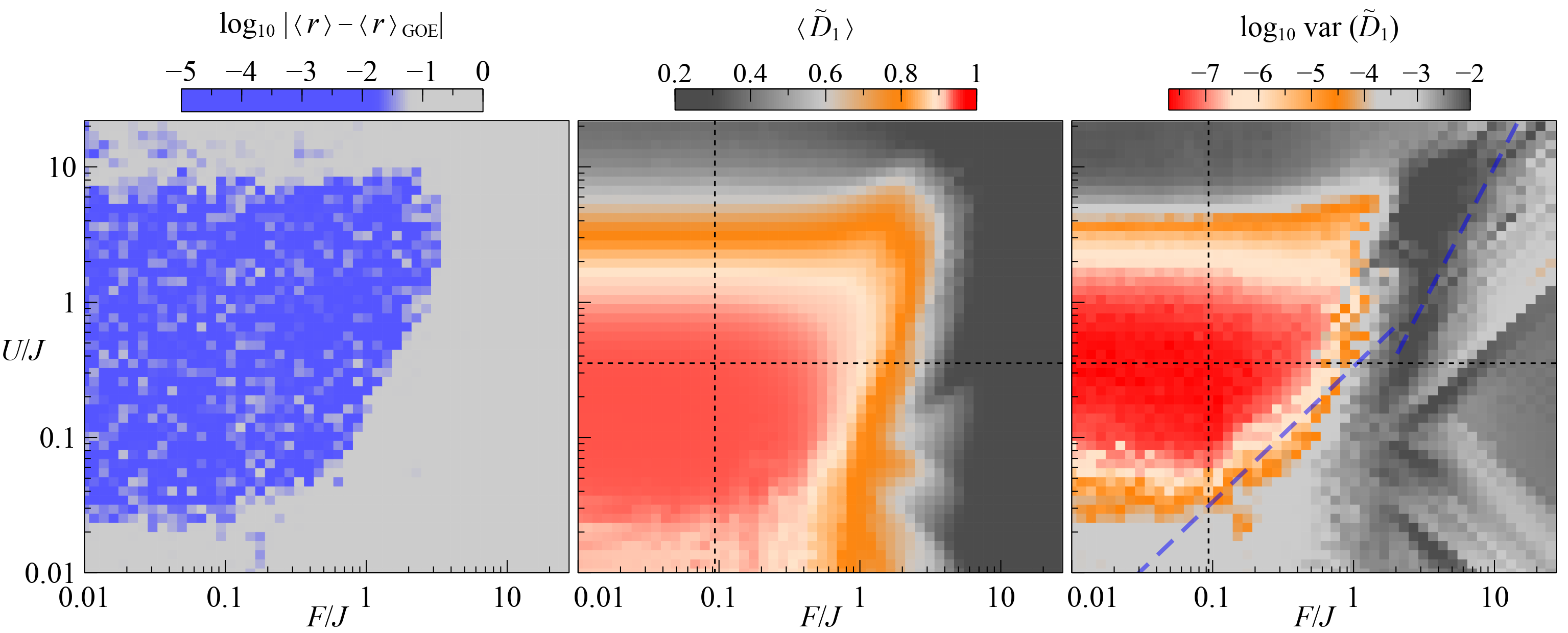}
	\caption{Chaotic phase for the homogeneous Fock state $\ket{\vec{1}}$ with $L=N=11$ ($\HSD=352\,716$) as a function of $U/J$ and $F/J$ revealed by (left) the mean level spacing ratio $\langle r \rangle$, (middle) the mean fractal dimension $\aDq{1}$, and (right) the variance $\vDq{1}$. All figures of merit are obtained from $\approx 200$ eigenstates with energies closest to $E=0$ for each pair $(U/J, F/J)$ ($50\times50$ points). Dotted lines highlight the trajectories $U/J=0.354$ and $F/J=0.0935$ considered in Fig.~\ref{fig:Scaling}. Blue dashed lines mark the critical line separating the regimes of unstable (left region) and stable (right region) Bloch oscillations obtained in Refs.~\cite{Kolovsky2009,Kolovsky2016a}.}
	\label{fig:DenPlotsE0L11}
\end{figure*}

The analysis presented in the previous section resolves the location and extension of the chaotic phase in the excitation spectrum as the model parameters are modified. Nonetheless, experimentally, the emergence of ergodicity is assessed from the dynamical behaviour of local (i.e., involving a reduced number of degrees of freedom) observables from initial density configurations which are typically Fock states in the onsite basis. Depending on the initial state considered, one follows a particular trajectory in the $\eps$ parametric space (as $J$, $F$ or $U$ are changed), which entirely determines whether and how the chaotic phase is traversed \cite{Pausch2025}. 

In this section, we characterize the chaotic phase from the perspective of the homogeneous Fock state at unit density,
\begin{equation}
 \ket{\vec{1}}\equiv\ket{1,1,\ldots,1}, 
 \label{eq:Fock1}
\end{equation}
which is typically considered in experiments with ultracold bosons \cite{Cheneau2012,Meinert2014a,Meinert2014b,Kaufman2016,Rispoli2019,Lukin2018,Bohrdt2020,Takasu2020,Leonard2023}. This state has energy 
\begin{equation}
 E_{\ket{\vec{1}}}\equiv \braket{\vec{1}|H|\vec{1}}=0,
 \label{eq:E1}
\end{equation}
for any value of the system parameters, due to our choice of tilt [see Eq.~\eqref{eq:BHH}]. 
A fixed value of the energy does not correspond to a fixed value of $\eps$, and the $E_{\ket{\vec{1}}}$ trajectory in the parametric space of the system is indicated by a solid black line in the $\langle r \rangle$ density plots of Figs.~\ref{fig:OverviewU1} and \ref{fig:OverviewJ1}. While for strong interactions the homogeneous state is close to being the ground state of the system ($\eps\to0$), for large tilt or large tunneling the spectrum becomes effectively symmetric around $E=0$, and hence $E_{\ket{\vec{1}}}$ approaches $\eps=0.5$ in those limits. 
The state's trajectory in Figs.~\ref{fig:OverviewU1} and \ref{fig:OverviewJ1} is accompanied by a shaded region which corresponds to the energy width $\pm\sigma_{\ket{\vec{1}}}$, obeying
\begin{align}
 \sigma_{\ket{\vec{1}}}^2 &= \braket{H^2}_{\ket{\vec{1}}} -\braket{H}_{\ket{\vec{1}}}^2 \notag \\ 
 &= 4J^2 (N-1).
 \label{eq:sigma1}
\end{align}
The latter provides the width of the local density of states (LDOS) of $H$ with respect to $\ket{\vec{1}}$, i.e., the energy range over which the Fock state has a noticeable overlap with the eigenstates of the system. 

By following the $E=0$ trajectory in the $(\eps,J/U)$ and $(\eps, F/J)$ spaces, one can see how the chaotic phase unfolds from the perspective of the state $\ket{\vec{1}}$, hence identifying the range of parameter values where ergodicity could be experimentally observed dynamically with such initial state. The chaotic phase for the homogeneous Fock state as a function of $U/J$ and $F/J$ is shown in Fig.~\ref{fig:DenPlotsE0L11} for a system with $L=N=11$, where the usual figures of merit are numerically obtained by considering the $\approx200$ eigenstates with energies closest to $E_{\ket{\vec{1}}}$. The ergodic domain is best recognizable from the behaviour of $\vDq{1}$, and roughly extends within the region $0.04\lesssim U/J\lesssim 4$ and $F/J\lesssim 1.5$, albeit the upper limit in $F$ increases with the interaction strength. It is also worth emphasizing that the observed chaotic phase lies within the region of unstable Bloch oscillation dynamics, identified from a mean-field approach in Refs.~\cite{Kolovsky2009,Kolovsky2016a}, and demarcated by a dashed blue line in Fig.~\ref{fig:DenPlotsE0L11}.

\begin{figure*}\centering
	\includegraphics[width=.95\columnwidth]{\figdir/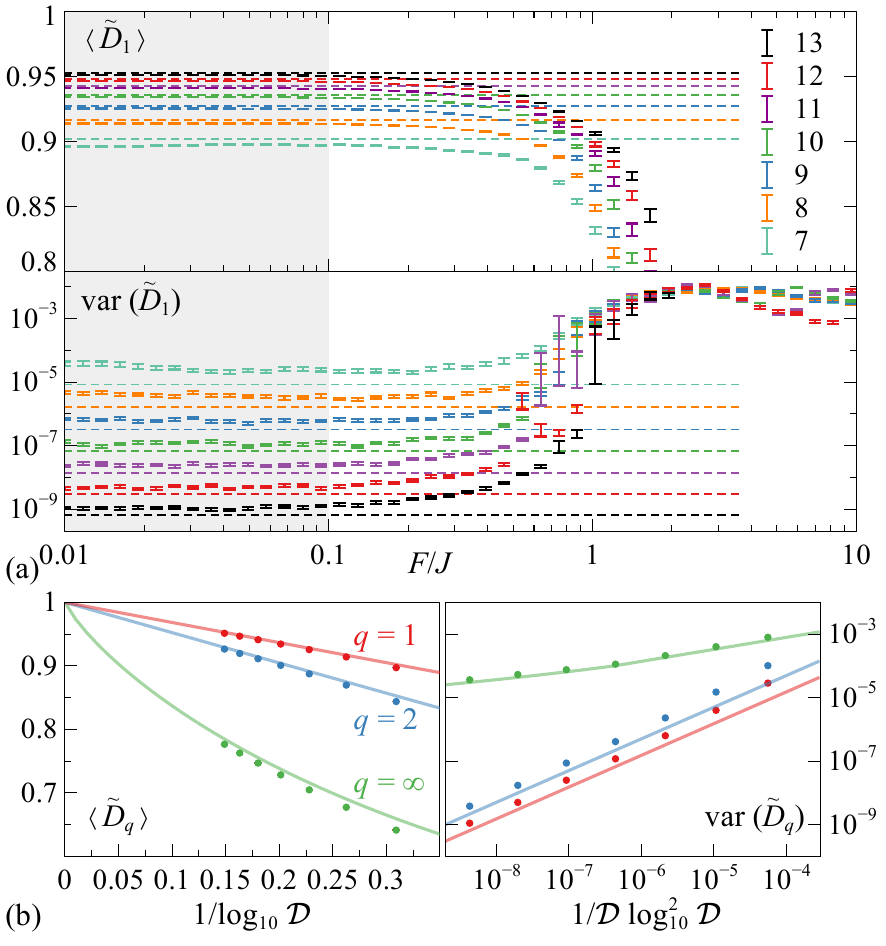}\hspace*{6mm}
	\includegraphics[width=.95\columnwidth]{\figdir/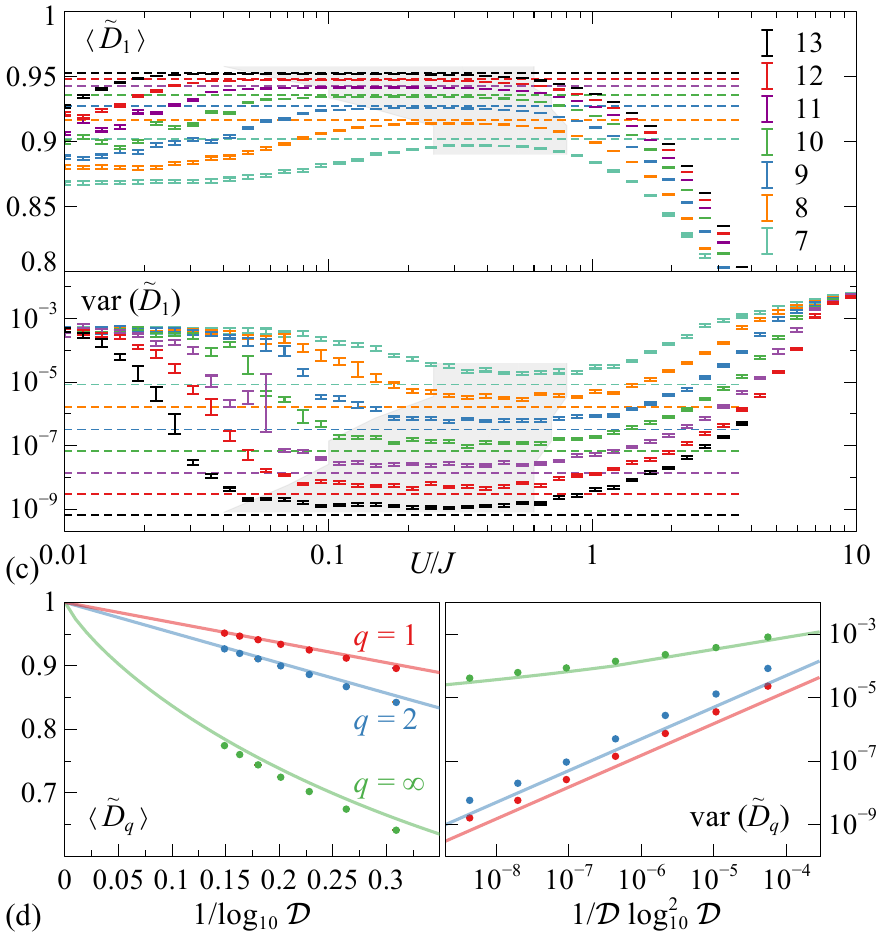}
	\caption{Development of the chaotic phase at $E=0$ for (left) fixed $U/J=0.354$ and (right) fixed $F/J=0.0935$ (see dotted trajectories in Fig.~\ref{fig:DenPlotsE0L11}). 
	Panels (a) and (c) show $\aDq{1}$ and $\vDq{1}$ versus $F/J$ and $U/J$, respectively, for varying system size $L\in[7,13]$ ($\mathcal{D}\in[1716,5\,200\,300]$). 
	Dashed lines correspond to GOE predictions. Panels (b) and (d) display the scaling of $\aDq{q}$ and $\vDq{q}$, for $q=1,2,\infty$, averaged over the deep chaotic regime [shaded regions in (a) and (c), respectively], versus Hilbert space dimension, where solid lines are GOE results. Errorbars are contained within symbol size whenever not visible.}
	\label{fig:Scaling}
\end{figure*}

In order to understand the dependence of the ergodic phase on system size at fixed density, we evaluate the mean and variance of $\Dq{1}$ for $L\in[7, 13]$ along the dashed trajectories indicated in Fig.~\ref{fig:DenPlotsE0L11}, corresponding to $U/J=0.354$ and $F/J=0.0935$ (cutting through the deep chaotic regime), and show the results in Figs.~\ref{fig:Scaling}(a) and \ref{fig:Scaling}(c), respectively. 
For fixed interaction strength, the position and extension of the chaotic phase as a function of the tilt remains fairly stable as system size grows: The value of $\Dq{1}$ rises and $\vDq{1}$ becomes ever smaller as $L$ increases for $F/J<1$, both comparing favourably with the GOE predictions, indicated by dashed lines, for the corresponding Hilbert space dimensions. On the other hand, for fixed tilt, the chaotic phase as a function of $U/J$ becomes clearly enhanced with system size: While the onset of chaos as $U/J$ decreases from the $J=0$ integrable limit exhibits a relatively weak $L$ dependence, the ergodic regime   extends to lower $U/J$ values the larger $L$, indicating that in the thermodynamic limit an arbitrarily small $U$ may be enough to trigger the emergence of the chaotic phase. This latter feature has also been observed in the absence of tilt \cite{Pausch2020}.

To provide a complete comparison against RMT, we show in Figs.~\ref{fig:Scaling}(b) and \ref{fig:Scaling}(d) the values of $\aDq{q}$ and $\vDq{q}$, for $q=1,2,\infty$, averaged over the deep chaotic regimes, highlighted respectively by the shaded regions in Fig.~\ref{fig:Scaling}(a) and \ref{fig:Scaling}(c), versus Hilbert space dimension. As can be observed, the numerical results follow closely the trends for GOE eigenvectors, indicated by solid lines [see Eqs.~\eqref{eq:D1goe} and \eqref{eq:vD1goe} and Ref.~\cite{Pausch2020}]. Hence, in the chaotic phase of the tilted Bose-Hubbard model, the dominant finite-size corrections of the convergence path toward ergodic eigenvectors in the thermodynamic limit exhibit the same functional dependence on $\HSD$ as those for RMT. 

\section{Conclusions}
\label{sec:conclusions}
We have characterized the chaotic phase of the tilted Bose-Hubbard system by analyzing its spectral and eigenvector features,  providing a valuable phase diagram for future theoretical and experimental studies of this system.
First, we analyze such features 
as energy-resolved functions of the tunneling strength $J/U$ for selected fixed values of the tilt $F/U$. Results show that the width of the chaotic region can be increased by a factor of three for moderate values of the external field, $F/U \lesssim 1$. From this point, the growth of the tilt strength leads to a shrinking of the chaotic regime. An analogous study by fixing the interaction strength reveals that the extension of the chaotic phase in terms of the tilt parametric range is maximized for $U/J\approx 1$. 


We check that the spectrum and eigenstate properties of the system in the chaotic regime are in accordance with the predictions for the Gaussian orthogonal ensemble of random matrix theory (RMT): We observe a mean value of the level spacing ratio $\langle r \rangle = 0.5307$,  and mean value of the first generalized fractal dimension approaching one, $\langle \tilde{D}_1 \rangle \to 1$. Moreover, the variance of $\tilde{D}_1$ over close-in-energy eigenstates exhibits a pronounced suppression (values tending to zero) within the chaotic domain, making this figure of merit a remarkably sensitive probe of quantum chaos.  
The chaotic phase is thus populated by ergodic eigenstates that share the same structure in Fock space. Additionally, the Kullback-Leibler divergence analysis for the $r$ distribution further confirms the unambiguous agreement with RMT whenever the chaotic phase emerges. 

To explore the possibility of observing ergodicity in a dynamics experiment, we keep track of the energy trajectory of the homogeneous Fock state at unit density, and see that the chaotic domain extends within the region $0.04\lesssim U/J\lesssim 4$ and $F/J\lesssim 1.5$ for a system with $11$ bosons. Upon increasing Hilbert space dimension up to $\mathcal{D}=5.2\times 10^6$ at fixed density, for fixed interaction strength, the position and width of the chaotic phase as a function of the tilt remains stable. On the other hand, for fixed tilt, the chaotic phase as a function of $U/J$ becomes enhanced with system size: The onset of chaos tends to lower $U/J$ values the larger the system size, indicating that, in the thermodynamic limit, and arbitrarily small interaction may be enough to induce ergodicity. The scaling also reveals that the dominant finite-size corrections of the convergence path toward ergodic eigenvectors have the same functional dependence on $\HSD$ as those for RMT.


\begin{acknowledgments}
We thank \'Oscar Due\~nas for a careful reading of the manuscript. 
The authors acknowledge support by Ministerio de Ciencia e Innovaci\'on/Agencia Estatal de Investigaci\'on MCIN/AEI (Spain) 
through Grant No.~PID2020-114830GB-I00. 
A.R.~acknowledges support by the German Research Foundation (DFG) through Grant No.~402552777.
This research has made use of the high performance computing resources of the Castilla y Le\'on Supercomputing Center (SCAYLE, www.scayle.es), financed by the
European Regional Development Fund (ERDF), and of the CSUC (Consorci de Serveis Universitaris de Catalunya) supercomputing resources. 
We thankfully acknowledge RES resources provided by the Galician Supercomputing Center (CESGA) in FinisTerrae III to activity FI-2024-2-0027.
The supercomputer FinisTerrae III and its permanent data storage system have been funded by the Spanish Ministry of Science and Innovation, the Galician Government and the European Regional Development Fund (ERDF).
\end{acknowledgments}
\section*{Data Availability Statement}
The data to reproduce all figures in this article are openly available at \href{http://hdl.handle.net/10366/164618}{http://hdl.handle.net/10366/164618}.

\bibliographystyle{apsrev4-2}
%
\end{document}